# Cell-Scale Dynamic Modeling of Membrane Interactions with Arbitrarily Shaped Particles

Didarul Ahasan Redwan[1], Justin Reicher[2], Xin Yong[1,2]*

[1] Department of Mechanical and Aerospace Engineering, University at Buffalo, Buffalo, NY 14260, USA.

[2] Department of Mechanical Engineering, Binghamton University, Binghamton, NY, 13902

## Abstract

Modeling membrane interactions with arbitrarily shaped colloidal particles, such as environmental micro- and nanoplastics, at the cell scale remains particularly challenging, owing to the complexity of particle geometries and the need to resolve fully coupled translational and rotational dynamics. Here, we present a force-based computational framework capable of capturing dynamic interactions between deformable lipid vesicles and rigid particles of irregular shapes. Both vesicle and particle surfaces are represented using triangulated meshes, and Langevin dynamics resolves membrane deformation alongside rigid-body particle motion. Adhesive interactions between the particle and membrane surfaces are modeled using two numerical schemes: a vertex-to-vertex mapping and a vertex-to-surface projection. The latter yields more accurate wrapping energetics, as demonstrated by benchmark comparisons against ideal spheres. The dynamic simulations reveal that lower particle-to-vesicle mass ratios facilitate frequent particle reorientation and complete membrane wrapping, while higher mass ratios limit orientation changes and stabilize partial wrapping. To illustrate the framework's versatility, we simulate interactions involving cubical, rod-like, bowl-shaped, and tetrahedral particles with spherical, cigar-shaped, or biconcave vesicles. This generalizable modeling approach enables predictive, cell-scale studies of membrane–particle interactions across a wide range of geometries, with applications in environmental biophysics and nanomedicine.

*Email: xinyong@buffalo.edu

## 1. Introduction

Marine plastic waste fragments into micro- and nanoplastics (MNPs), broadly defined as particles with diameters $\leq 5$ mm. Once released into the environment, these particles can enter a variety of cell types through endocytosis, micropinocytosis, and phagocytosis, potentially triggering harmful biological responses.[1, 2] Among these pathways, phagocytosis plays a central role in the uptake of larger particles, where the plasma membrane actively deforms to engulf extracellular materials into an intracellular vesicle.[1-4] Unlike engineered colloids that are predominantly spherical, MNPs often display highly irregular and anisotropic shapes.[5, 6] These morphological features, along with surface chemistry,[7-11] size,[12, 13] and deformability,[14-16] critically influence how particles interact with and are internalized by cellular membranes.[17, 18]

Anisotropic particle shapes introduce curvature heterogeneity, resulting in membrane wrapping behaviors that differ fundamentally from those of spherical particles.[12, 18-27] Computational and theoretical studies have demonstrated that for non-spherical particles, the local variation in curvature leads to spatially heterogeneous bending energy costs. This heterogeneity can stabilize partially wrapped states and significantly influence vesicle morphology.[19, 20, 22, 23, 28] For example, Dasgupta et al.[22] reported that partially wrapped ellipsoids can remain stable because the membrane resists bending around the high-curvature tip of the particle. Other simulations have reported a two-stage sequence for ellipsoid particles: an initial flat-side adhesion, followed by a tip-first reorientation that lowers the overall bending penalty and enables complete engulfment. In such cases, the particle's large aspect ratio amplifies the energy barrier to reach complete wrapping.[19, 21, 24, 29]

Experimental observations support these computational findings. A recent study by van der Ham et al.[25] confirmed orientation-dependent wrapping behavior, showing that a long, rod-shaped particle initially adopts a "surfing" state along the surface of a giant unilamellar vesicle (GUV) before becoming engulfed. During this process, the particle undergoes further reorientations prior to complete uptake. Similarly, Azadbakht et al.[30] showed that variations in the neck curvature of dumbbell-shaped particles modulate both the speed and completeness of engulfment by GUV. In particular, the constricted neck region imposes a kinetic barrier, rendering wrapping highly sensitive to membrane tension. Together, these studies underscore that shape anisotropy not only introduces local curvature variation but also governs the sequence of particle reorientation events necessary for complete membrane envelopment.[21, 24, 25, 29, 31]

In biological cells, particle–membrane interactions can involve more complex physics. While the wrapping process during phagocytosis can be adequately predicted by passive membrane elasticity and adhesion energetics,[19, 20, 22, 23, 32, 33] living cells frequently employ active processes to drive membrane deformation and facilitate the uptake of exogenous substances beyond what passive interactions alone can achieve. Protrusive forces from actin polymerization and contractile forces generated by myosin motors actively remodel the membrane to promote particle internalization.[4, 34-39] Furthermore, membrane proteins, such as BAR-domain scaffolds and curvature-sensitive complexes, facilitate uptake dynamics by locally inducing curvature and recruiting cytoskeletal elements.[40-46] While recognizing the biological importance of these active membrane-deformation mechanisms, our study intentionally employs a simplified modeling framework that excludes active forces and cytoskeletal coupling. This passive model enables precise quantitative comparisons with controlled experiments involving synthetic vesicles or cells lacking active endocytic machinery, thus providing a foundational understanding that can guide future incorporation of active mechanisms.

The choice of numerical method for passive wrapping simulations depends on the specific objectives of the study and the particular aspects of membrane dynamics of interest (see Table S1 in the ESI). Energy-based simulations using Helfrich bending theory and numerical methods such as Monte Carlo predict wrapping phase diagrams and equilibrium vesicle shapes by minimizing membrane bending and adhesion energies.[4, 23, 28, 44] However, these thermodynamic approaches fall short of revealing the kinetic pathway of wrapping. Specifically, they yield only equilibrium configurations and do not resolve the time-dependent dynamics of particle–vesicle interactions. This limitation is particularly significant for anisotropic and irregular shapes, whose orientation and dynamic responses to membrane interactions critically influence cellular uptake, as emphasized by experimental studies. Addressing these kinetic pathways thus requires a dynamic simulation framework capable of computing membrane-mediated forces and iteratively updating the translation and rotation of irregularly shaped particles

We extend our previously developed force-based model[47] to simulate the interaction dynamics of vesicles with anisotropic particles, including those lacking analytical shape representations. Within a unified modeling framework, both deformable membranes and arbitrarily shaped rigid particles are represented using triangulated surface meshes. Two distinct numerical schemes are implemented to model adhesion between particle and membrane surfaces, and their accuracy in predicting wrapping energetics is systematically compared. Furthermore, we characterize the role of particle inertia in translation and rotational dynamics, explicitly capturing the force–torque coupling mediated by the membrane. Finally, we demonstrate the versatility of our approach by simulating interactions between non-spherical vesicles and particles with complex geometric features, including significant curvature variations and convex–concave transitions. This framework offers a versatile tool for exploring membrane–particle interactions beyond idealized geometries, supporting future research in both environmental biophysics and nanomedicine, particularly regarding shape-dependent uptake mechanisms at the cellular scale.

## 2. Methods

### 2.1. Triangulated surface model for vesicle

The process by which a vesicle membrane wraps around a particle is governed by the competition between bending energy $E_{bend}$ and adhesion energy $E_{adh}$. The bending energy captures the cost of deforming the membrane, which is described by the Canham-Helfrich (CH) energy functional. This model is integral to predicting membrane deformation under various physical conditions and is described as follows[48, 49]

$$E_{bend} = \oint dA\,[2\kappa_b(H - H_0)^2 + \kappa_b G] \quad (1)$$

Here, $H$, $G$, and $H_0$ represent the mean, Gaussian, and spontaneous curvatures of the membrane, respectively. Bending moduli $\kappa_b$ and Gaussian curvature moduli $\kappa_g$ characterize the strength of the membrane against different deformation modes. This study does not consider vesicle topology changes and thus the Gaussian curvature term is neglected according to the Gauss-Bonnet theorem.

In addition to the bending energy, a vesicle model must include area and volume constraints to maintain its mechanical stability and realistic membrane behavior. The area constraint energy ensures that the vesicle's surface area remains constant, reflecting the lateral incompressibility of the lipid bilayer (the per area lipid varies little without extreme tension).[50] Similarly, the volume constraint energy penalizes variations in internal fluid volume, thereby maintaining the osmotic pressure difference between the vesicle and its surroundings.[33] These constraints work together to control the vesicle shape under physiologically relevant conditions, such as variations in membrane tension or osmotic concentrations. We employed simple quadratic forms for these two energies, given by

$$E_a = \kappa_a \frac{(A - A_t)^2}{A} \quad (2)$$

$$E_v = \kappa_v \frac{(V - V_t)^2}{V} \quad (3)$$

Here, $\kappa_a$ is the area expansion modulus, while $A_t$ and $A$ denote the preferred and current surface area of the vesicle membrane, respectively. In a similar fashion, $V_t$ and $V$ represent the target and current vesicle volume, respectively. $\kappa_v$ acts as a modulus that controls the strength of osmotic pressure imbalance when the vesicle is deformed from its natural shape by particle interaction.[33]

The membrane shape corresponding to the lowest energy state can be theoretically derived by varying the CH energy functional. However, solving the resulting shape equation is often intractable due to complex geometric attributes. To overcome this, the membrane is modeled as a discretized two-dimensional surface using a triangulated mesh, which enables the application of discrete differential geometry to evaluate the membrane energy functional. Specifically, the membrane surface $S$ is represented by a mesh network composed of $N_v$ vertices, $N_e$ edges, and $N_f$ triangular faces. Each vertex $v_i$ corresponds to a point on the surface embedded in a three-dimensional laboratory frame. All vertices collectively define the global shape of the membrane. A triplet of connected vertices defines each triangular face. The numerical derivations and formulations of geometric variables are detailed in the following references.[48, 51, 52] Details of vesicle mesh generation and energy calculations for the discretized membrane can be found in our previous work.[47]

**2.2. Triangulated surface representation of arbitrarily shaped particles**

Analytical representations are often inadequate for particles with complex geometries, such as those exhibiting widely varying curvature, concave regions, and sharp edges or corners. Such irregular features typically cannot be captured by simple closed-form expressions.[53] Due to these limitations, triangulated surface meshes (similar to the vesicle surface) provide a flexible solution. By approximating a surface with numerous small facets, a mesh can closely fit virtually any shape and even accommodate complex topographies with intricate surface curvature variations.[54] Five particles with diverse shapes are simulated using triangulated meshes in this study. Spherical particles are generated by three-level icosahedron subdivision to achieve near-uniform triangulation. To enable direct comparisons with previous simulations and experiments, cube- and rod-like particles are generated using a MATLAB mesh generator[55] according to the equations of superellipsoids,[20, 23] $x^6 + y^6 + z^6 = R_p^6$ with $R_p = 0.2$ and $[(x^2 + y^2)/a^2]^{n/2} + (z/b)^n = 1$, with $n = 6$, $a = 0.15$, and $b = 0.3$, respectively. We also simulate cubical particles with rounded corners through a Minkowski sum of a cube and a sphere with a cube edge length of 0.6 and a rounded edge/corner radius of 0.2. The mesh for a bowl-shaped particle is created based on an initial oblate spheroid mesh with a major axis of 0.33 and a minor axis of 0.24. We carve out a smooth concave “bowl” region on its lower hemisphere by displacing the $z$ positions of the vertices with $z \leq 0$ following $z_{new} = z + D_{max} \cos\left(\pi \sqrt{x^2 + y^2} / 2R_{bowl}\right)$. Here, $D_{max}$ is the dimple depth which is set to $0.27$ and $R_{bowl} = 0.33$. For a tetrahedral particle, we generate an ideal tetrahedral mesh using an edge length 0.6 and then smooth the corners and edges using the libigl mesh library.[56]

**2.3. Mesh-based schemes for defining membrane–particle interaction**

The membrane–particle interaction energy can be written as the surface integral

$$E_{adh} = \int V\left(d(x)\right) dA \tag{4}$$

Here, $d(x)$ is the separation distance between the membrane at point $x$ and the particle surface. $V(d)$ is the interaction potential density acting between the two interacting surfaces. As the membrane surface is discretized into triangular faces, the total energy is computed by summing the contributions from the interacting vertices, weighted by their associated Voronoi areas $A_i$[32, 57]

$$E_{adh} = -\sum_i V(d_i) \cdot A_i \quad (5)$$

Here, $d_i$ is the distance from the membrane vertex $v_i$ to the particle surface. The adhesion between membrane and particle is modeled using Morse potential, taking the following form[58]

$$V(d_i) = U\left[e^{(-2d_i/\rho)} - 2e^{(-d_i/\rho)}\right] \quad (6)$$

In this expression, non-negative $U$ is the adhesion strength, and positive $\rho$ defines the interaction range. This potential has a minimum value of $-U$ at $d_i = 0$, indicating attraction that lowers the total free energy (see Figure 1). To reduce the computational cost, we introduce a numerical cutoff $r_c$ (on the order of potential range) and include only those membrane vertices for which $d_i < r_c$ in the calculation. We also define a dimensionless adhesion strength $u = UR_p^2/\kappa_b$ for spherical particles, where $R_p$ is the radius. In the case of a non-spherical particle, we replace $R_p$ with an effective radius $R_{eff} = \sqrt{A_p/4\pi}$ obtained from the total particle mesh surface area $A_p$. This $R_{eff}$ is then used to rescale the adhesion strength, yielding $u_{mod} = UR_{eff}^2/\kappa_b$ for arbitrarily shaped particles.

We determine $d_i$ for an arbitrarily shaped particle represented by a triangulated mesh using two alternative geometric approaches. In the vertex-to-vertex scheme, each membrane vertex is paired with the nearest vertex on the particle mesh (a nearest-neighbor mapping) to estimate $d_i$. In the vertex-to-surface scheme, $d_i$ is computed as the shortest Euclidean distance from the membrane vertex to any triangular face of the particle mesh. While the two approaches define the local membrane–particle separation differently for complex particle shapes, they allow us to impose adhesion between the particle and membrane, which drives wrapping and uptake. The implementation details of both methods are described as follows.

***Vertex-to-vertex adhesion scheme*** To compute the adhesion energy between the vesicle and particle surface, the Morse potential is applied based on the pairwise distance between the vertices of the two triangulated surfaces. This approach, referred to later as the vertex-to-vertex scheme, draws inspiration from prior studies on vesicle–vesicle aggregation in fluid flows.[59] This method establishes virtual bonds between nearest neighboring pairs of vesicle and membrane vertices (see Figure 2), maintaining a one-to-one correspondence. The resulting bond distances are then used to calculate the adhesion energy and force. Notably, we ensure that bonds are formed only between mutually nearest vertex pairs, thereby avoiding redundant or overlapping interactions. The detailed algorithm for the vertex-to-vertex model is presented in the Appendix.

***Vertex-to-surface adhesion scheme*** The vertex-to-surface scheme computes the shortest

Euclidean distance from a given membrane vertex to the triangulated surface of the particle.[60, 61] This approach measures the true minimal separation between two triangulated surfaces, independent of mesh resolution or vertex alignment. It ensures that even if the closest approach occurs between a vertex and a face (or edge) of the other mesh (rather than between two vertices), the distance is accurately captured.

Specifically, we describe how to find each vertex's closest point on the target surface to obtain the unsigned distance.[56] Consider a vesicle vertex $v_i$ at the position $\mathbf{P}$ and a closed triangulated surface $S^1$ (see Figure 3), $P$ is orthogonally projected onto the plane of a triangle. If the projection point $Q$ resides within the triangle (determined via barycentric coordinates or a cross-product test), then the normal distance $|\mathbf{P}-\mathbf{Q}|$ is a candidate. If $Q$ lies outside the triangle, we instead project $P$ onto each of the three edges of the triangle, clamping the projection to the segment endpoints. Thus, the closest point is either this edge projection or the nearer vertex. The Euclidean distance from $P$ to each of these boundary candidates for a triangle is computed and the smallest is retained. After evaluating all triangles, the global minimum of these candidate distances yields the closest point $C$ on $S^1$ and the corresponding triangle, and the unsigned distance is simply $|\mathbf{P}-\mathbf{C}|$.

As the negative distance branch of the Morse potential is important for avoiding the particle unphysically penetrating the membrane, we define the sign of the distance between a vesicle vertex $v_i$ and the particle surface $S^1$. This is achieved by taking the dot product of an angle-weighted pseudomonal[62] $\mathbf{n}_\alpha$ with the vector from the surface to the point, $\mathbf{P}-\mathbf{C}$. The sign of the dot product indicates the positional relationship: $\mathbf{n}_\alpha \cdot (\mathbf{P}-\mathbf{C}) > 0$ if $P$ lies outside the surface $S^1$; $\mathbf{n}_\alpha \cdot (\mathbf{P}-\mathbf{C}) < 0$ if $P$ lies inside $S^1$; $\mathbf{n}_\alpha \cdot (\mathbf{P}-\mathbf{C}) = 0$ if $P$ lies exactly on $S^1$. This criterion works regardless of whether the closest point $C$ is on a face, an edge, or coincides with a vertex of the mesh,[62] reliably determining the sign of the distance even at sharp features where a normal is not uniquely defined.

### 2.3. Time integration scheme

To simulate the motion of the vesicle membrane, the velocity-Verlet algorithm has been chosen as the time integration scheme, consistent with approaches used in molecular dynamics packages such as LAMMPS.[63] The motion of each vertex on the triangulated vesicle surface is governed by a deterministic form of the Langevin equation:[64, 65]

$$m\frac{d\mathbf{r}_i^2}{dt^2} = \mathbf{F}_i^{tot} - \gamma\frac{d\mathbf{r}_i}{dt} \tag{7}$$

where $m$ is the vertex mass. The total deterministic force $\mathbf{F}_i^{tot}$ includes contributions from bending elasticity, area and volume constraints, and membrane-particle adhesion, with the detailed calculations described in Ref.[47]. The damping term with the drag coefficient $\gamma$ serves two primary purposes: it ensures the numerical stability of the integrator and approximates viscous dissipation from the surrounding fluid. This approach is commonly adopted in simulations of particle–membrane interactions at the cellular scale.[66, 67]

It is well documented that hydrodynamic drag on particles increases with confinement, such as proximity to a membrane or solid boundary. While Stokes' law describes constant drag in unbounded fluids, the presence of nearby surfaces results in higher and position-dependent drag.[68, 69] However, such variation primarily affects the absolute time scale of particle motion, rather than the mechanistic sequence of orientational transitions or the equilibrium wrapping states.[66] Accounting for position-dependent drag induced by the presence of the membrane would require either explicit solvent modeling (e.g., using dissipative particle dynamics or lattice Boltzmann methods) or analytical treatment of hydrodynamic interactions.[70] These approaches significantly increase computational cost and are beyond the scope of the present work. Moreover, many mesoscopic simulations of nanoparticle wrapping studies have similarly employed a constant damping coefficient.[64, 65] Therefore, we believe that the assumption of a constant friction coefficient is justified for capturing the role of particle inertia and adhesion strength in wrapping dynamics. We set the vertex mass and drag coefficient ($m = \gamma = 1$ in the simulation units) to model an intermediate damping regime.

Notably, this fully deterministic Langevin formulation[66] omits stochastic thermal noise, solvent hydrodynamics (beyond simple drag), and active cellular factors such as cortical tension and receptor–cytoskeleton coupling. This simplification allows the adhesion–bending energy balance to be treated in a computational and analytically tractable manner. However, we acknowledge that this approach has limited quantitative predictive power at the nanoscale, where thermal fluctuations on the order of $k_\mathrm{B}T$ (with $k_\mathrm{B}$ being the Boltzmann constant), hydrodynamic dissipation, and active forces can facilitate energy barrier crossing and influence uptake kinetics.

To explore the coupled dynamics of the vesicle–particle system, the translational and rotational motion of nanoparticle is explicitly modeled using rigid body dynamics. A similar Langevin dynamics also governs the translation motion of the particle's center of mass, consistent with the vesicle vertex dynamics. The equation is given by

$$m_p \frac{d\mathbf{r}_{com}^2}{dt^2} = \sum\nolimits_j \mathbf{F}_j^p - \gamma_p \frac{d\mathbf{r}_{com}}{dt} \tag{8}$$

Here, $m_p$ is the total mass of the rigid particle, $\mathbf{r}_{com}$ is the particle's center of mass position, and $\gamma_p$ is the friction coefficient accounting for translational damping. $\mathbf{F}_j^p$ is the reaction force to membrane adhesion acting on the particle surface. For each membrane vertex $v_i$ experiencing a force $\mathbf{F}_i^{adh}$, we apply $\mathbf{F}_j^p = -\mathbf{F}_i^{adh}$ at the corresponding point $j$ on the particle, which is either the nearest mesh vertex (Figure 2) in the vertex-to-vertex scheme or the closest surface point $C$ (Figure 3) in the vertex-to-surface scheme. Thus, $\sum_j \mathbf{F}_j^p$ gives the net force driving particle translation, while the drag term damps its motion. We fix $m_p/\gamma_p = 1$ to ensure a particle damping regime consistent with the membrane vertices.

The rotational dynamics is governed by the total torque generated from off-center reactions applied to the particle surface. The torque about the particle's center of mass is calculated as

$$\boldsymbol{\tau} = \sum_j (\mathbf{r}_j - \mathbf{r}_{com}) \times \mathbf{F}_j^p \tag{9}$$

In addition to the torque arising from the reaction force to membrane adhesion, we incorporate an effective rotational drag torque to model viscous dissipation acting on the particle. The rotational drag torque is proportional to the particle's angular velocity in the body frame and is scaled by its principal moments of inertia.[63] The drag torque is calculated in the body frame as

$$\boldsymbol{\tau}_d^b = -\tilde{\gamma}_{rot}\left[I_{xx}\omega_x^b, I_{yy}\omega_y^b, I_{zz}\omega_z^b\right] \tag{10}$$

Here, $\tilde{\gamma}_{rot}$ is the rotational damping coefficient. $I_{xx}$, $I_{yy}$, and $I_{zz}$ are the principal moments of the inertia tensor $\mathrm{I}$, while $\omega_x^b$, $\omega_y^b$, and $\omega_z^b$ are the components of the angular velocity vector in the body frame. This body-frame drag torque is then transformed into the laboratory frame using the rotation matrix $\mathrm{R}$ associated with the particle's current orientation

$$\boldsymbol{\tau}_d^s = \mathrm{R}\boldsymbol{\tau}_d^b \tag{11}$$

This laboratory-frame drag torque is added to the total torque balance when updating the particle's rotational dynamics, providing essential rotational damping similar to linear damping applied to the particle's center of mass and vesicle vertices. Such damping prevents any artificial or ongoing spinning that may arise from numerical noise or integration drift and is crucial for achieving physically meaningful equilibrium states.

Subsequently, the angular momentum $\mathbf{L}$ and angular velocity $\boldsymbol{\omega}$ are calculated from the total applied drag ($\boldsymbol{\tau}_{tot} = \boldsymbol{\tau} - \boldsymbol{\tau}_d^s$) on the particle over a time step $\Delta t$

$$\mathbf{L} = \boldsymbol{\tau}_{tot}\Delta t \tag{12}$$

$$\boldsymbol{\omega} = \mathrm{I}^{-1}\mathbf{L} \tag{13}$$

The orientation is tracked using a quaternion $q = q_0 + q_1\mathbf{i} + q_2\mathbf{j} + q_3\mathbf{k}$, which is updated through

$$\frac{dq}{dt} = \frac{1}{2} q \otimes \boldsymbol{\omega} \tag{14}$$

where $\otimes$ denotes the quaternion product. The updated quaternion is then normalized, followed by the reconstruction of the body frame to obtain new positions of all particle vertices. This quaternion-based formulation avoids singularity in rotation kinematics and ensures stable, smooth integration of rotational motion.[63, 71, 72] The detailed simulation parameters are given in the following table.

A mapping between the simulation units and physical units can be established by referencing comparable GUV experiments. The characteristic length scale is set by the vesicle radius $l_0 = 10\ \mu m$, reflecting typical GUV size. The characteristic time scale is $\tau = 0.05\ s$, based on experimentally observed uptake time.[25] The energy unit $E_0$ is defined by the bending modulus $\kappa_b$ of typical phospholipid membranes, as bending energy represents the dominant energetic contribution during vesicle deformation. With this unit system, all other simulation parameters, including the area expansion modulus, the volume constraint modulus, membrane and particle masses, damping coefficients, and interaction potentials, can be consistently converted to physical units via dimensional analysis (Table 1). Notably, the resulting membrane vertex drag coefficient closely matches the estimate given by Stokes' law for water.

**Table 1: Standard simulation parameters**

| Parameters | Numerical Values | Physical Values |
| --- | --- | --- |
| Bending modulus ($\kappa_b$) | 0.01 [51] | $20\ k_B T$ [73, 74] |
| Area expansion modulus ($\kappa_a$) | 1.0 | $8.3 \times 10^{-5}\ nN \cdot \mu m^{-1}$ |
| Volume constraint modulus ($\kappa_v$) | 2.0 (for controlled-volume) [51]<br>0.0 (for uncontrolled-volume) | $1.6 \times 10^{-5}\ nN \cdot \mu m^{-2}$ |
| Spontaneous curvature ($H_0$) | 0 | $0\ \mu m^{-1}$ |
| Vesicle surface area ($A_t$) | 4π | $1.3 \times 10^3\ \mu m^2$ |
| Morse potential range ($\rho$) | 0.01 | $100\ \mu m$ |
| Simulation time step ($\Delta t$) | 0.001 | $5 \times 10^{-5}\ s$ |
| Vesicle mass ($m_v$) | 10242 | $2 \times 10^{-10}\ kg$ |
| Membrane vertex drag coefficient ($\gamma$) | 1.0 | $4.1 \times 10^{-6}\ nN \cdot s \cdot \mu m^{-1}$ |
| Particle mass ($m_p$) | 1.0 | $2 \times 10^{-14}\ kg$ |
| Particle translational drag coefficient ($\gamma_p$) | 1.0 | $4.1 \times 10^{-6}\ nN \cdot s \cdot \mu m^{-1}$ |

| Particle rotational damping coefficient ($\tilde{\gamma}_{rot}$) | 1.0 | $20\ \mathrm{s}^{-1}$ |
|---|---|---|

## 3. Results and Discussion

### 3.1. Comparison of interaction schemes for meshed particles

To evaluate how different numerical schemes affect the particle–membrane interaction, we characterize the energy profile of a spherical vesicle wrapping an external or internal nanosphere of radius $R_p = 0.3$ with reduced adhesion strength $u = 2$. Both the particle and vesicle meshes are generated by successive subdivisions of an icosahedron, yielding 2,562 vesicle vertices ($N_v$) and 602 particle vertices ($N_p$). In these simulations, the particle is held stationary and the wrapping energy is quantified when a steady-state configuration is reached. A harmonic umbrella potential is employed to sample the states with high wrapping fractions, which are naturally not attainable.[47] The wrapping fraction $\chi$ is calculated from the ratio of current adhesion energy to the theoretical value corresponding to full wrapping, $\chi = E_{adh}/UA_p$. The results of the meshed sphere using the vertex-to-vertex and vertex-to-surface scheme are compared against the data in our previous study modeling a parametric sphere of the same size.[47] Notably, no volume constraint is applied for consistency.

As shown in Figure 4, the vertex-to-vertex scheme clearly overestimates both bending and total energies compared to the energetics predicted by the parametric sphere. This significant discrepancy is attributed to the calculation of surface distance. As the two triangulated surfaces approach each other, the vertex-to-vertex bonds shorten. However, as the membrane mesh locally conforms to the particle mesh, the computed bond lengths cannot reduce to zero even when membrane vertices are brought onto the triangular faces of the particle mesh. This residual distance results in the evaluation of the Morse potential and adhesion forces at artificially larger distances, leading to inaccurate energy profiles. This effect is more pronounced when there is a high local curvature generated at high $\chi$. Regardless of whether particles are inside or outside the vesicle, the total energy sharply increases when the wrapping fraction exceeds 0.6. The neck formation at this high wrapping fraction introduces very large local curvature at the membrane–particle interface, causing the vertex-to-vertex distance to deviate increasingly from true surface distance. The exact threshold at which this diverging behavior occurs depends on the resolution of the membrane and particle meshes. Finer meshes may extend this limit somewhat, but the qualitative constraint remains.

In contrast, the vertex-to-surface scheme renders accurate adhesion forces, even at higher wrapping fractions where local membrane curvature also varies largely. The predicted energy curves closely align with those of the parametric sphere. Such an accurate evaluation of surface distance is essential for capturing contact mechanics and wrapping dynamics in our simulations.

Due to this markedly improved agreement, the vertex-to-surface scheme is adopted for all subsequent simulations.

To further benchmark our numerical framework, we conducted simulations replicating the system in Yu et al.,[23] who reported normalized bending energy as a function of the wrapping fraction for a cube-like particle of size $R_p = 0.2$. In Figure S1, we compare their results with those from our current model for two distinct cases: (i) a stationary particle with all translational and rotational motion suppressed, and (ii) a freely moving particle with both degrees of freedom enabled. Our model shows good agreement with Yu et al.'s data at low and high wrapping fractions. However, for intermediate wrapping states $(\chi \sim 0.3 - 0.7)$, our simulations predict a lower equilibrium bending energy.

This discrepancy arises from fundamental differences in simulation methodology. In Yu et al.'s study, the membrane region adhering to the particle was held fixed, and thus the vesicle shape evolved with a fixed contact line. In contrast, our approach dynamically evolves the membrane and particle meshes independently, without constraining the adhesion interface/boundary. In other words, given a certain wrapping fraction, the membrane in our model can adaptively cover different regions of particles of the same area. This relative mobility between the meshes allows the system to explore energetically more efficient wrapping pathways, which contributes to the lower bending energies observed in our simulations at intermediate wrapping degrees.

### 3.2. Effect of particle inertia on interaction dynamics

In our Langevin-dynamics framework, the particle-to-vesicle mass ratio $m_p/m_v$ (where $m_v = mN_v$ is the total vesicle mass) governs the particle's inertia relative to the interacting membrane. This affects force–torque coupling, particularly how applied torques translate into particle rotation. Heavier particles tend to retain their initial orientation, while lighter ones respond more readily to vesicle deformation. To examine this effect, we simulate cubical particles with three mass ratios—0.0001, 0.0005, and 0.001—interacting with a biconcave vesicle with its reduced volume constrained at $0.65$ (defined as the ratio of the vesicle volume to that of a sphere having the same surface area).[75, 76] A rescaled adhesion strength $u_{mod} = 4.0$ is applied enable a robust wrapping transition. The particle is initially oriented with one corner contacting the vesicle's north pole ( "corner-attack" configuration).

As shown in Figure 5, the two lighter particles ($m_p/m_v = 0.0001$ and 0.0005) undergo multiple reorientations during wrapping, while the heaviest particle ($m_p/m_v = 0.001$) retains its original orientation throughout. The lightest particle exhibits the most dynamic behavior, rapidly transitioning to a "face-attack" configuration ( Figure 5a). Because the top dimple of the biconcave vesicle is intrinsically shallow, this orientation smooths local curvature variations, reducing bending energy. However, as wrapping progresses, further advancement in the face-attack configuration requires membrane bending around multiple corners, which becomes energetically unfavorable. Around $t \sim 500$, the particle reverts to the corner-attack orientation to leverage the dimple's natural concavity to reduce bending costs. Later, at $t \sim 800$, it returns to the face-attack pose, enabling full wrapping of side faces, which is maintained thereafter (see Supplementary Video 1). These rapid transitions reflect the particle's low inertia.

The intermediate-mass particle $(m_p/m_v = 0.0005)$ initially remains in the corner-attack configuration and transitions only once to the face-attack orientation later in the simulation (Figure 5b). This delayed reorientation suggests that inertia suppresses the early-stage transition observed for the lightest particle, but becomes less influential as wrapping progresses and adhesion forces dominate. In contrast, the heaviest cubical particle $(m_p/m_v = 0.001)$ shows no significant reorientation and remains in the corner-attack configuration throughout (Figure 5c). Its high inertia prevents the transition to a face-attack pose, limiting membrane wrapping. As shown in Figure S2, the wrapping fraction for the heaviest particle plateaus around 0.55, whereas the lighter particles reach values near 0.8. This behavior aligns with recent experimental observations of poly(l-lactic-co-glycolic) acid nanoparticle uptake by macrophages, where heavier particles were internalized more slowly, with longer kinetic half-times and frequent arrest in partially wrapped states.[77] Overall, these results demonstrate how particle inertia influences rotational dynamics and wrapping efficiency through force–torque interactions with a deformable membrane. Given our focus on resolving rigid body dynamics, we use the lightest particle $m_p/m_v = 0.0001$ in subsequent simulations. The effects of particle and vesicle mesh densities are studied for this mass ratio. The wrapping kinetics shows convergent behavior when particle mesh density is higher than $N_p = 702$ (Figure S3a). When vesicle density is changed, the overall wrapping pathway and final configuration remain the same but the kinetics shows variations (see Figure S3b). This difference is due to the bond density and formation sequence between the particle and membrane surfaces being altered for different mesh densities.

### 3.3. Cubical and rod-like particles interacting with a spherical vesicle

We examine the interaction between a cubical particle and a spherical vesicle to understand further how adhesion strength influences the final wrapping states of anisotropic particles. A series of simulations is performed with increasing adhesion strength, ranging from $u_{mod} = 2.0$ to $10.0$, as shown in Figure 6a. At low adhesion strength $(u_{mod} = 2.0)$, the particle retains its initial orientation with only weak membrane interaction and minimal wrapping. As the adhesion strength increases to intermediate values, the cubical particle reorients into the face-attack configuration, thereby increasing the effective contact area and lowering local adhesion energy. Interestingly, at higher adhesion strengths $(u_{mod} = 8.0 - 10.0)$, the particle becomes fully wrapped by the vesicle while reverting to the corner-attack orientation. These transitions between corner-attack and face-attack configurations are consistent with the dynamics previously observed in Figure 5.

We extend our analysis to rod-like particles introduced in two distinct initial orientations: side-wise (lying horizontally along the vesicle surface) and tip-wise (aligned vertically). The interaction dynamics obtained from our simulations are compared with experimental observations reported from van der Ham et al.[25] (Figure S4), showing excellent agreement. For side-wise particles, the membrane readily wraps upon initial contact, with the rods remaining horizontal in the partially wrapped state. As the membrane contact advances and begins to cover the edge of the rod tip, the particle gradually pivots upward along its long axis, eventually achieving full engulfment (Figure S3a, b). This gradual reorientation is accompanied by steady increases in both membrane bending energy and wrapping fraction at $u_{mod} \geq 6.0$ (Figure S5a, c), indicating entry into the engulfment phase at $u_{mod} = 6.0$. The vesicle's reduced volume decreases concurrently (Figure S4b), reflecting internal accommodation of the particle.

For tip-wise particles, the initial wrapping proceeds more slowly due to the small contact area and high energy cost associated with bending around the tip (Figure S5c). Rapid engulfment begins once the particle breaks symmetry and transitions into a slanted orientation (Figure S4c, d). Compared to side-wise particles, this slanted entry leads to sharper increases in bending energy, reduced volume, and wrapping fraction observed at $u_{mod} \geq 8.0$.

Figure 6b,c illustrates how adhesion strength influences the final configuration. Side-wise particles achieve full wrapping at lower adhesion strengths ($u_{mod} = 6.0$) compared to tip-wise particles, which require $u_{mod} = 8.0$. At lower adhesion strengths, particles remain partially wrapped while largely retaining their initial orientations. At high adhesion strengths ($u_{mod} = 8.0$ to $10.0$), the vertical alignment of the particle and rotational symmetry of the vesicle are no longer maintained. This rapid change in curvature at the small catenoidal neck region is also reflected in the bending energy curves for high adhesion strengths. Together, these results demonstrate that the final orientation and wrapping state of anisotropic particles are governed by a combined effect of particle shape and adhesion strength.

### 3.4. Cubical particle interaction dynamics with a cigar-shaped vesicle

Capturing the wrapping dynamics is key to understanding particle uptake pathways in vesicles and cells. We examine the uptake of cubical particles by a cigar-shaped vesicle, focusing on morphological evolution across a range of rescaled adhesion strengths ($u_{mod} = 2.0 - 10.0$). The vesicle is modeled with a volume constraint corresponding to a reduced volume of 0.7,[75, 76] resembling tubular endosomes and Weibel–Palade bodies—elongated organelles that support long-range cargo transport and targeted release.[78, 79] Two initial contact locations are considered: the saddle-shaped waist with negative Gaussian curvature, and the highly convex north pole (Figure 7). In both cases, the particle begins in a corner-attack orientation for consistency.

Figure 7 illustrates how adhesion strength and initial contact location affect uptake. Particles contacting the waist achieve full wrapping at lower adhesion strength ($u_{mod} = 6.0$) compared to pole-attached particles, which require $u_{mod} = 10.0$. This difference is attributed to the fact that the saddle-shaped waist promotes membrane bending more readily than the convex pole. For waist-attached particles, wrapping proceeds through three stages: an initial lag phase with low wrapping fraction, a growth phase marked by reorientation to edge or face contact, and a final burst phase with rapid engulfment. These stages correspond to sharp increases in both bending energy and wrapping fraction at $u_{mod} \geq 6.0$ (Figure S4). The reduced volume initially decreases to accommodate the particle, then partially recovers due to the volume constraint, resulting in a capsule-like final shape (Video S4).

In contrast, pole-attached particles remain partially wrapped at lower adhesion strengths, arrested in the lag phase for $u_{mod} \leq 4.0$. At $u_{mod} = 6.0$ to $8.0$, only one face attaches to the pole region (Figure 7d), and full engulfment is observed only at $u_{mod} = 10.0$. During this process, the particle reorients from its initial corner-attack pose to a face-attack configuration, then returns to corner-attack after complete engulfment (Video S5). Substantial increases in bending energy are seen only at $u_{mod} = 10.0$. Correspondingly, reduced volume drops significantly without recovery, indicating a more energetically costly membrane deformation compared to waist interaction (Figure S5). These results show that adhesion strength, initial contact geometry, and local membrane curvature together dictate the efficiency, dynamics, and morphology of particles wrapping by non-spherical vesicles.

### 3.5. Interaction between bowl-shaped and tetrahedral particles with a biconcave vesicle

To further test the robustness of our computational framework, we simulate the wrapping of two geometrically complex particles: a bowl-shaped particle with opposing concave and convex surfaces, and a tetrahedral particle with sharp facets and pronounced curvature variation. A biconcave vesicle is used in both cases to provide a membrane with regions of both positive and negative curvature, mimicking red blood cell geometry.

In the first case, the bowl-shaped particle is placed with its rim contacting the vesicle surface (Figure 8a). Despite a high adhesion strength ($u_{mod} = 10.0$), wrapping remains shallow initially due to curvature mismatch and the associated bending energy cost. Over time, the particle shifts into the vesicle's dimple, where local curvature enables greater contact. Notably, the membrane opposite the adhering region bulges outward, forming a convex shape—consistent with previous observations of vesicle remodeling during spherical particle uptake.[47] When the particle is oriented with its concave face against the vesicle's convex waist (Figure 8b), complementary curvature allows full coverage of the concave surface with minimal membrane deformation. However, further wrapping around the particle rim is hindered by the steep bending penalty required to curve around the rim, even at $u_{mod} = 10.0$.

The final simulation features a tetrahedral particle positioned near the flatter side of the vesicle (Figure 8c). At $u_{mod} = 10.0$, full wrapping occurs spontaneously. The particle first approaches with a corner, then reorients to align a flat face for greater contact. As wrapping continues, it flips into an inverted pose that allows three faces to be enclosed, ultimately achieving full engulfment. Further rotation of the particle inside the vesicle follows. These simulations demonstrate the framework's capability to resolve complex wrapping behaviors driven by particle shape, adhesion strength, and membrane curvature.

## 4. Conclusion

In this work, we developed a force-based, dynamic simulation framework to resolve the translational and rotational dynamics of arbitrarily shaped particles interacting with fluidic vesicles. Both the deformable membrane and rigid particles are modeled using triangulated surface meshes. The method integrates the CH bending energy functional with a deterministic Langevin equation to describe the deformation of vesicles and the rigid body motion of particles. Adhesive interactions between the triangulated meshes are implemented via two numerical schemes: vertex-to-vertex (nearest-neighbor bonds) and vertex-to-surface (closest-point projections). While the vertex-to-vertex scheme performs well for modeling vesicle aggregation[59] or weak particle adhesion,[80-82] we demonstrated that the more accurate distance calculations in the vertex-to-surface scheme are essential for simulating substantial membrane wrapping during particle uptake.

Dynamic simulations of a cubical particle interacting with a biconcave vesicle show that higher particle inertia impedes the reorientations necessary for full engulfment. In contrast, lighter particles undergo multiple reorientations to achieve complete uptake. Our simulations of rod–vesicle interactions with different initial configurations reveal sequential orientational transitions closely matching the experimental observations. Overall, our results demonstrate that membrane wrapping is governed by the interplay between particle orientation, local membrane curvature, and adhesion strength. Particles adapt their orientations during uptake to reduce energetic barriers and promote favorable membrane deformation. When curvature compatibility is high, this adaptive behavior leads to complete engulfment; otherwise, wrapping is arrested in partial or metastable states. These principles hold across a range of particle and vesicle shapes, underscoring general mechanisms of anisotropic particle uptake. To conclude, this geometry-agnostic framework not only captures steady-state membrane deformations and energy landscapes but also resolves the time-resolved trajectory of particle entry. This work provides a versatile tool for studying environmental colloidal particles and guiding the design of anisotropic nanocarriers in biomedical applications at cellular scales.

## Author contributions

D. A. R., J. R., and X. Y. performed research; D. A. R. and J. R. analyzed data; and D. A. R. and X. Y. wrote the manuscript.

**Conflicts of Interest**

There are no conflicts to declare.

**Acknowledgment**

X. Y. gratefully acknowledges funding from the National Science Foundation for supporting this work through award 2034855/2448213. Computing time was provided by the Center for Computational Research at the University at Buffalo and the Theory and Computation facility of the Center for Functional Nanomaterials (CFN), which is a U.S. Department of Energy Office of Science User Facility, at Brookhaven National Laboratory under Contract No. DE-SC0012704. We would also like to acknowledge helpful discussions with Emad Pirhadi.

**Appendix**

**Algorithm: Vertex-to-vertex bond identification**

Input: $V_v$ (set of vesicle mesh vertices, size $N$), $V_p$ (set of particle mesh vertices, size $M$)

Output: <u>bonds</u> (list of mutual nearest-neighbor vertex index pairs)

Initialize $nearestParticle$ and $nearestVesicle$ to store nearest neighbor vertex indices

**for** each vesicle vertex $v_i \in V_v$ **do**:

    Compute the shortest distance for $v_i$, $d_i = \min_{v_j \in V_p} |\mathbf{v}_i - \mathbf{v}_j|$

    Store the index of the particle vertex corresponding to $d_i$, $nearestParticle[i] = j$

**end for**

**for** each particle vertex $v_j \in V_p$ **do**:

    Computer the shortest distance for $v_j$ as $d_j = \min_{v_i \in V_v} |\mathbf{v}_i - \mathbf{v}_j|$

    Store the index of the particle vertex corresponding to $d_j$, $nearestVesicle[j] = i$

**end for**

Initialize empty list bonds

**for** each vesicle vertex index $i = 1\ to\ N$ **do**:

    **if** the pair is mutually nearest, $nearestVesicle(nearestParticle[i]) == i$

**then** Append $(i, nearestParticle[i])$ to bonds

Return bonds

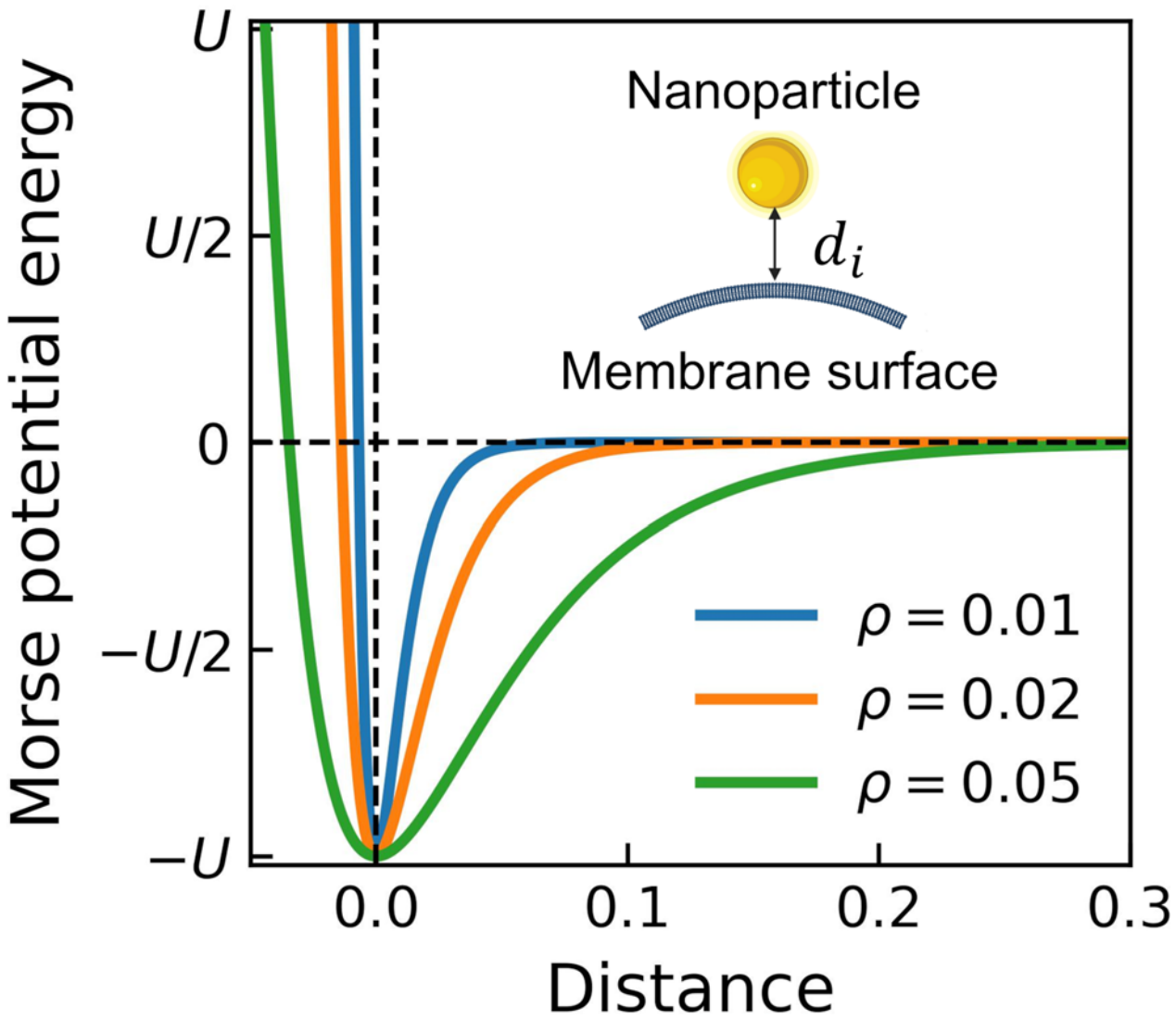

Figure 1: Nanoparticle–vesicle interaction potential. Morse potential profiles show how the interaction range is controlled by $\rho$. The inset schematically depicts a nanoparticle (yellow) approaching the vesicle membrane (gray) with the interaction distance $d_i$ defined between the particle and membrane surfaces.

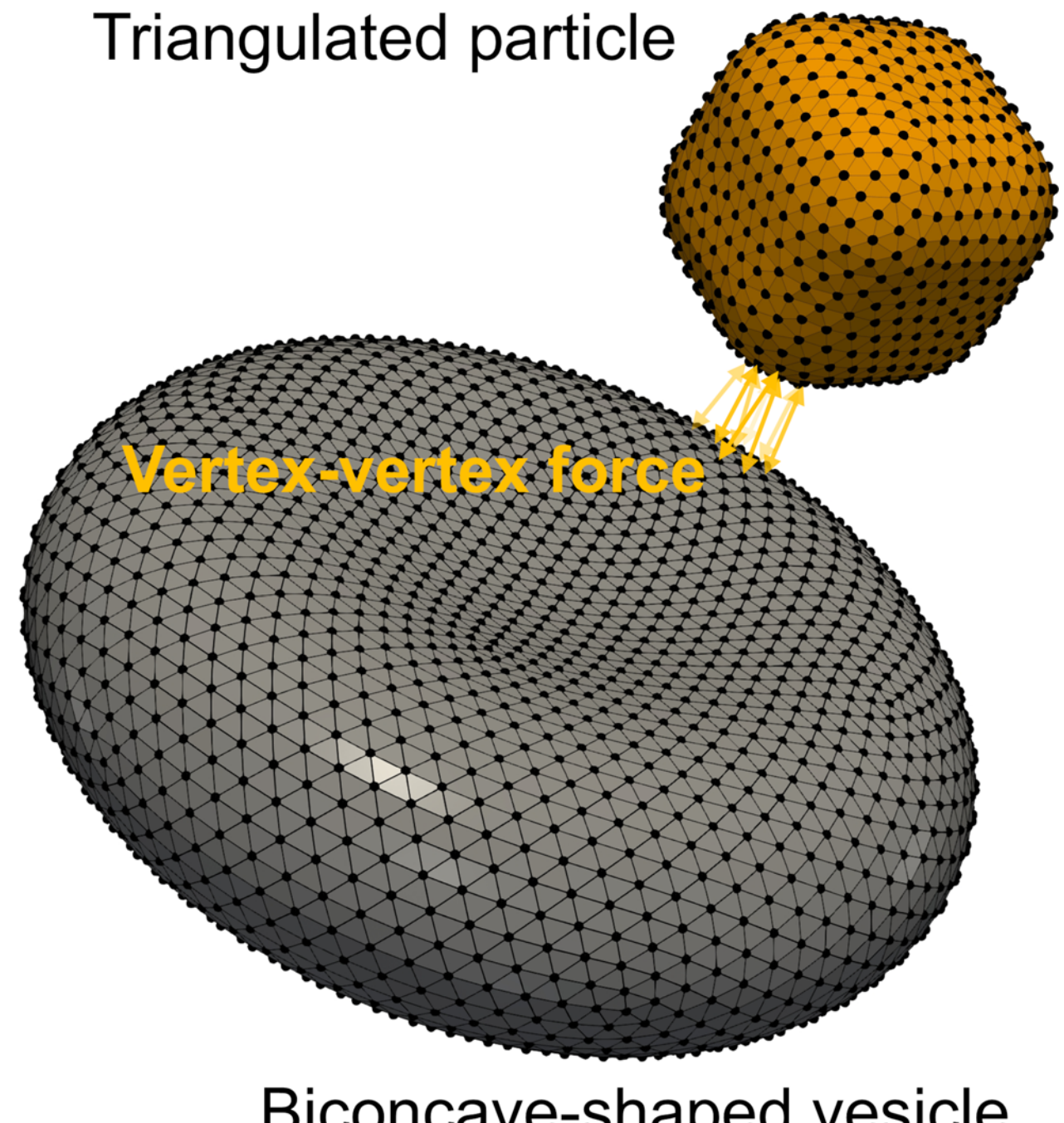

Figure 2: Vertex-to-vertex interaction scheme. The schematic illustrates the vertex-to-vertex scheme for calculating the interaction between a triangulated rigid particle (orange mesh) and a biconcave vesicle (gray mesh). Each yellow arrow represents a one-to-one "bond" formed between a membrane vertex and its nearest neighbor on the particle mesh. For visual clarity, the vesicle and particle are not shown at their relative physical sizes.

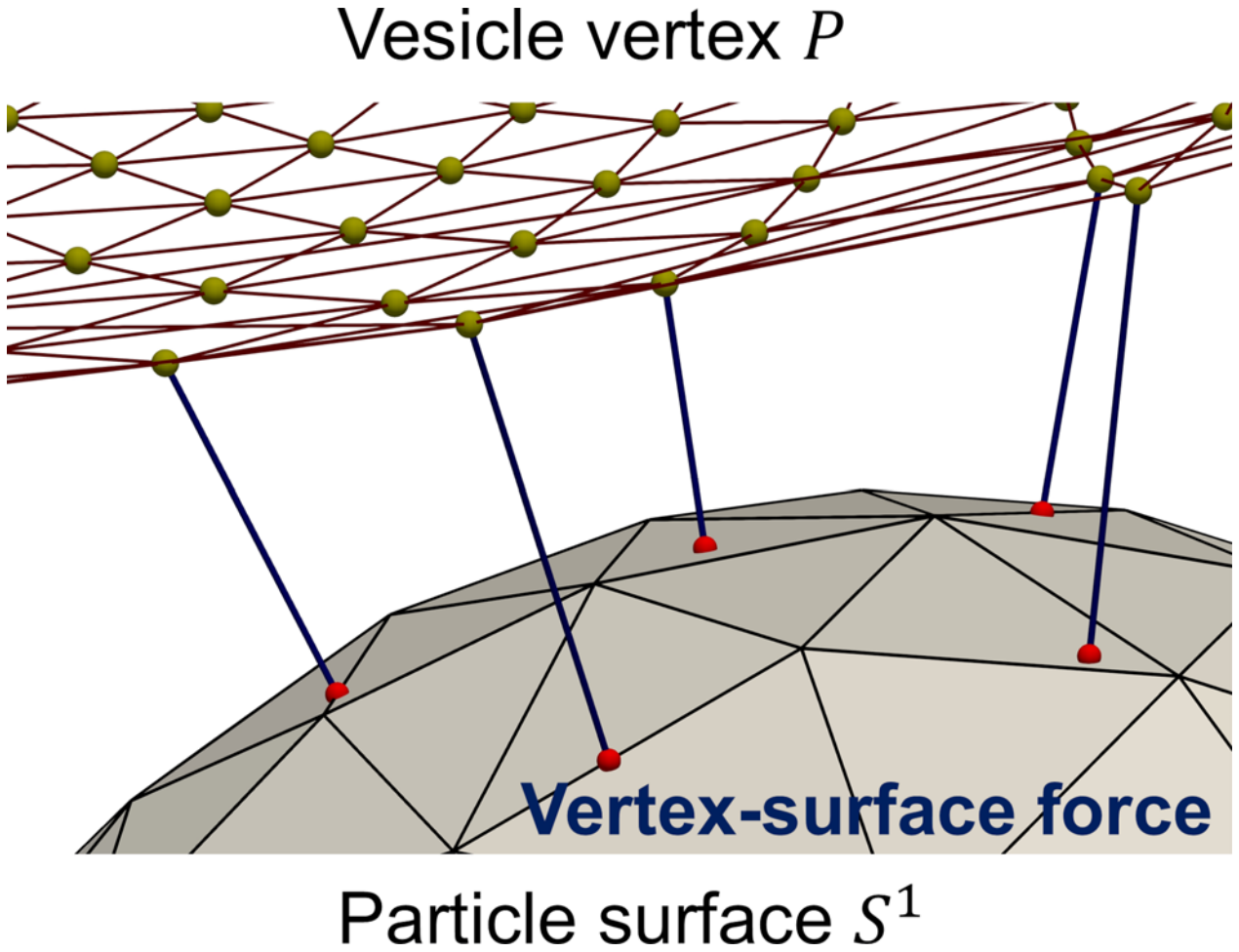

Figure 3: Vertex-to-surface interaction scheme. For each membrane vertex $P$ (yellow sphere), the closest point $C$ (red sphere) on the particle mesh is identified by minimizing $|\mathbf{P}-\mathbf{C}|$. The interaction force is calculated based on the shortest distance $|\mathbf{P}-\mathbf{C}|$, acting along the line connecting $P$ and $C$.

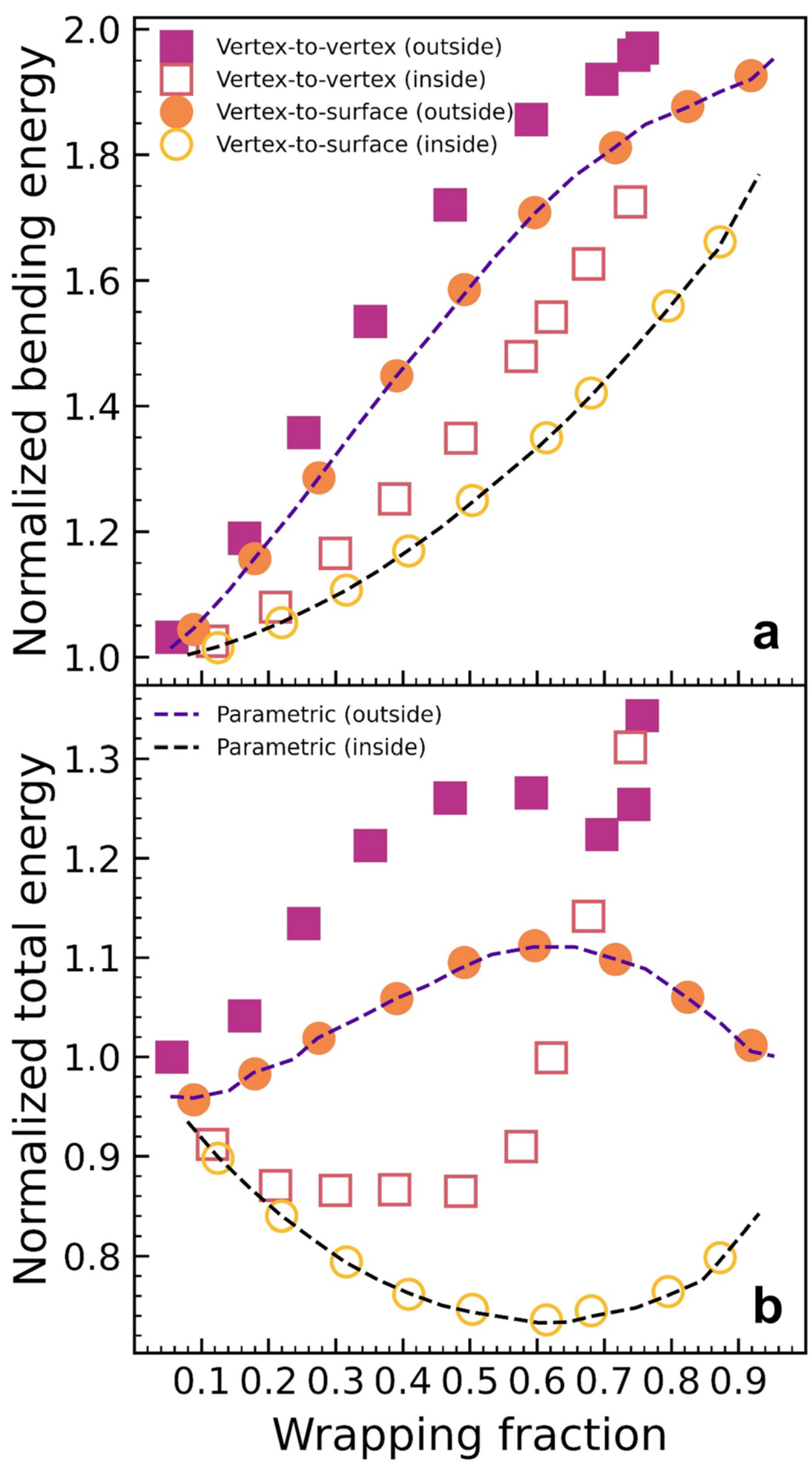

Figure 4: Quantitative comparison between the two interaction schemes for triangulated rigid particles. Equilibrium (a) bending and (b) total energies of a spherical vesicle interacting with a spherical particle of dimensionless radius of 0.3, reduced adhesion strength $u = 2.0$, and interaction potential range $\rho = 0.01$, plotted against the wrapping fraction $\chi$ of the particle. Different branches correspond to particles positioned inside and outside the vesicle. Dashed lines represent results for an ideal sphere from Ref. 47, while square and circular symbols denote simulations using triangulated spheres with vertex-to-vertex and vertex-to-surface interaction schemes, respectively.

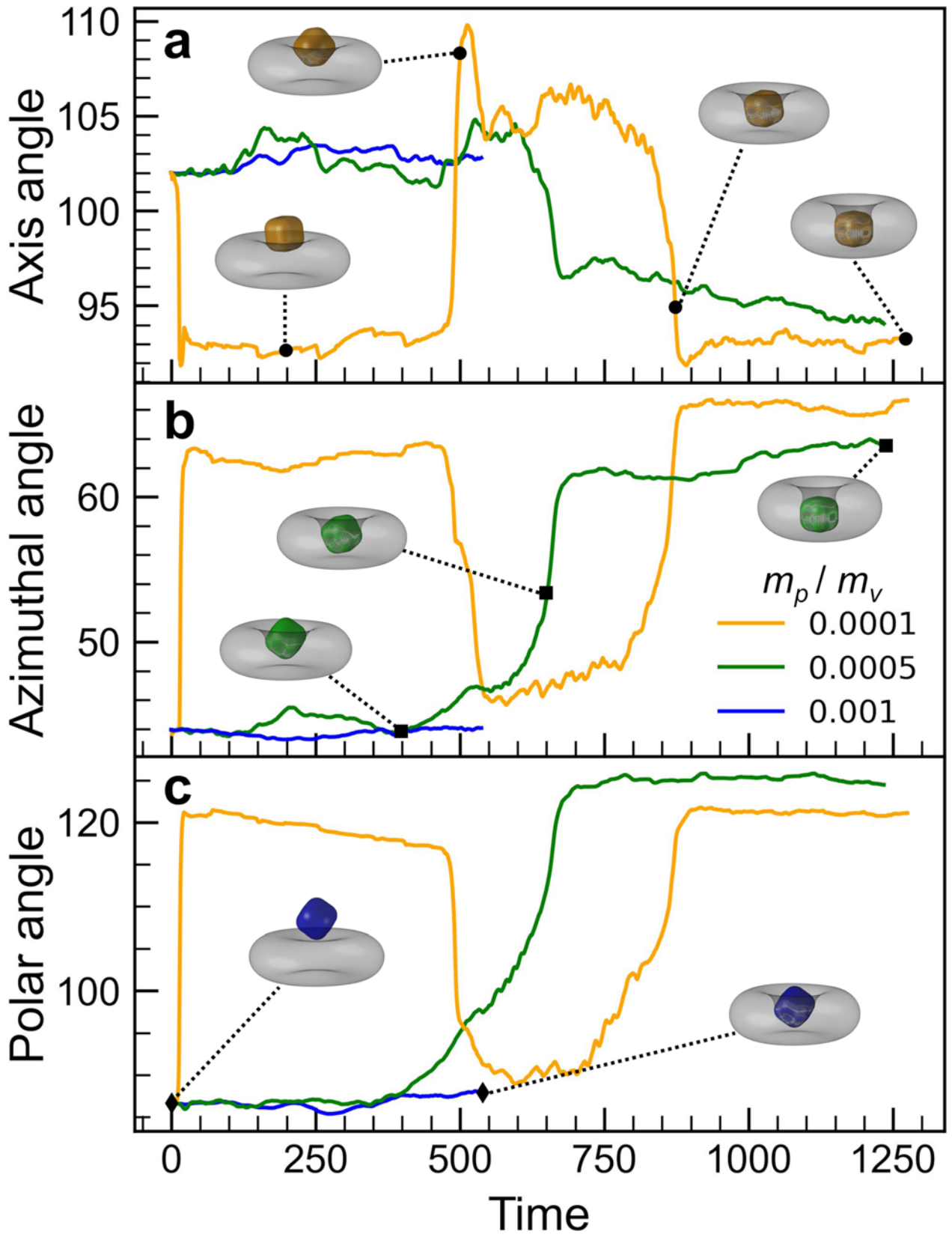

Figure 5: Mass-dependent rotational dynamics of rounded cubical particles ($u_{mod} = 4.0$) interacting with a biconcave vesicle. The particle's instantaneous orientation is described by the rotation from the laboratory frame to the particle's body frame, defined by the principal axes of its moment of inertia tensor. This rotation is parameterized using the Euler vector $\theta\mathbf{e}$, where the unit vector $\mathbf{e}$ indicates the direction of an axis of rotation and positive $\theta$ is the rotation angle following the right-hand rule. The direction of $\mathbf{e}$ is further expressed by its azimuthal and polar angles in the laboratory frame. Time evolutions of (a) rotation angle, (b) azimuthal angle, and (c) polar angle are shown for particles with particle-to-vesicle mass ratios $m_p/m_v = 0.0001$ (yellow), 0.0005 (green), and 0.001 (blue). Insets show corresponding vesicle–particle configurations at selected time points.

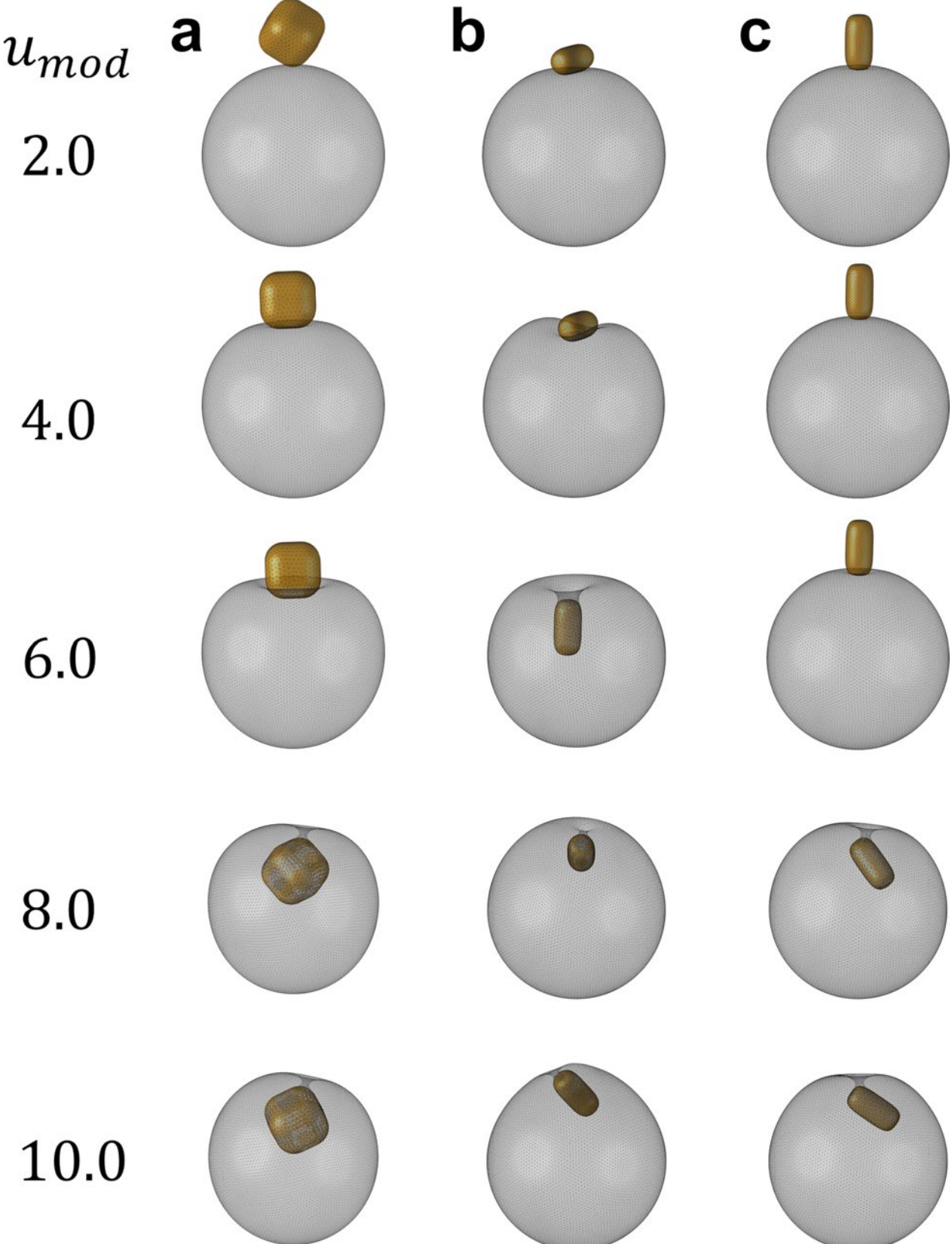

Figure 6: Effect of adhesion strength on anisotropic particles interacting with a spherical vesicle without the volume constraint. Final configurations with particles having different initial orientations at different rescaled adhesion strengths: (a) cubical particle with the corner contact, (b) rod-like particle with side-wise contact, and (c) rod-like particles with tip-wise contact. The side-wise rod achieves full wrapping at $u_{mod} = 6.0$, whereas both the cubical particle and tip-wise rod require $u_{mod} = 8.0$ for complete engulfment.

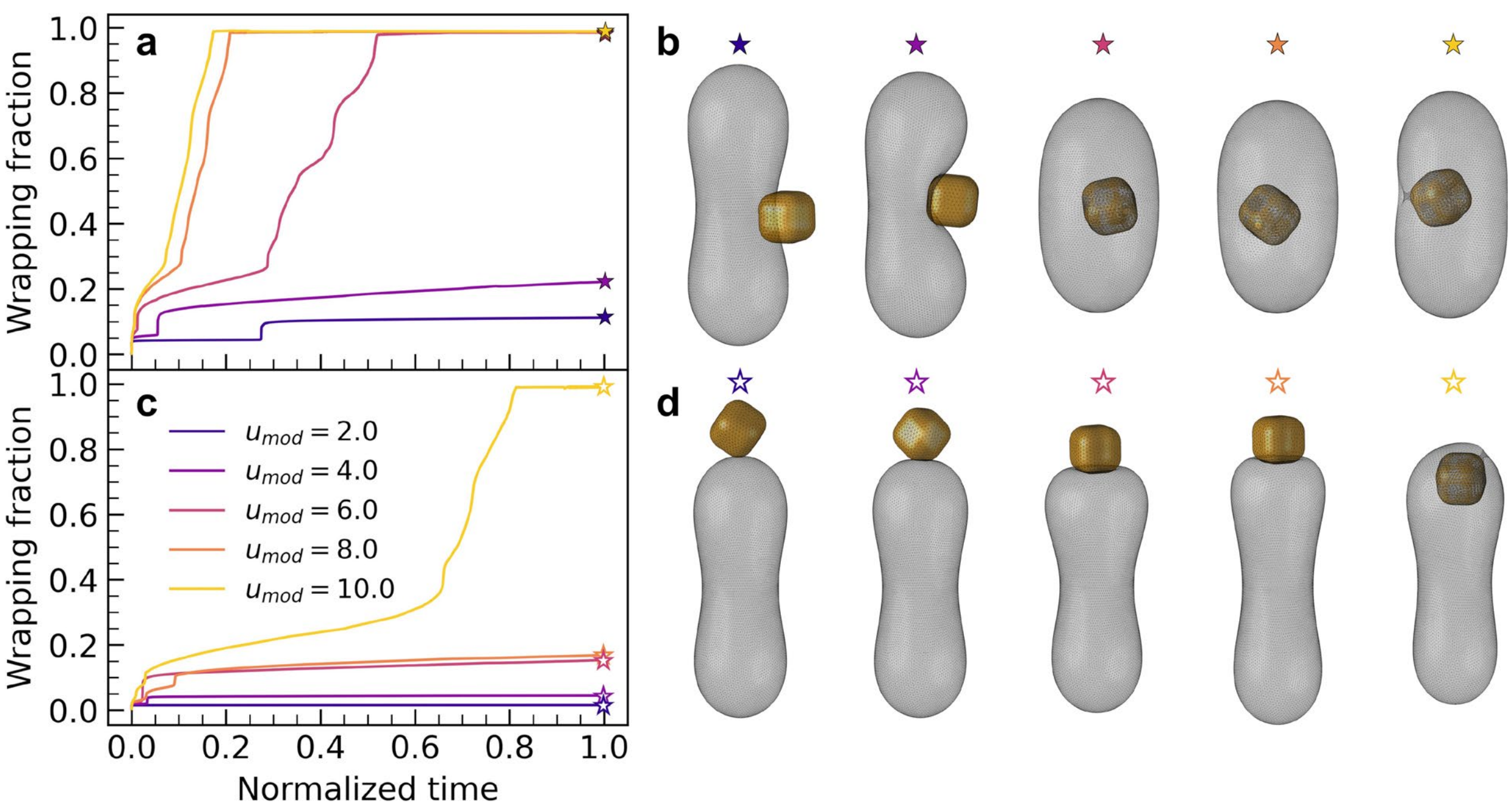

Figure 7: Interaction dynamics of cubical particles with a cigar-shaped vesicle. Time evolution of the wrapping fraction for a cubical particle contacting the (a) saddle-shaped waist and (c) convex pole of the vesicle at different rescaled adhesion strengths. Time is normalized by the maximum duration required to reach the steady-state configuration, which depends on the rescaled adhesion strength. Filled and open star markers indicate the final states for contact at the waist and pole, respectively. Corresponding simulation snapshots are shown in panels (b) and (d).

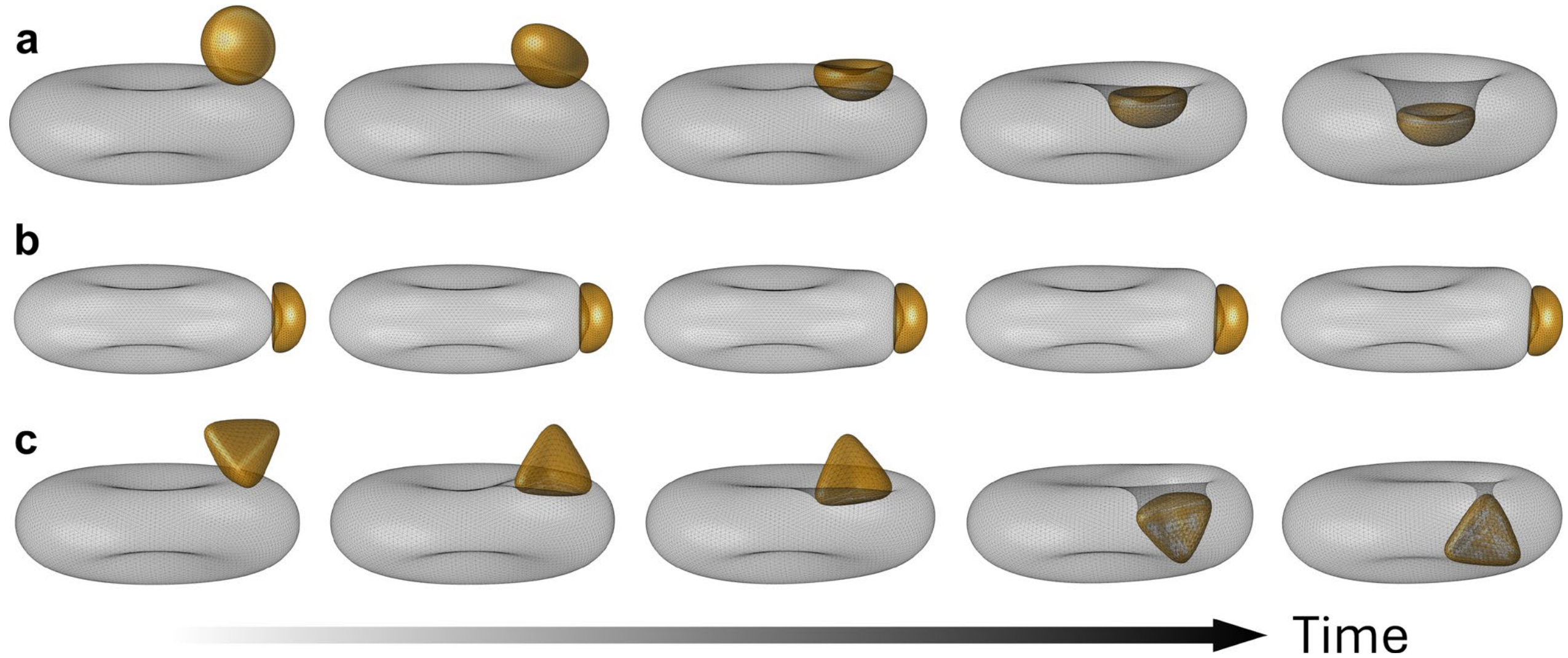

Figure 8: Biconcave vesicle interaction with bowl-shaped and tetrahedral particles. (a) Bowl-shaped particle ($u_{mod} = 6.0$) initially contacting the vesicle with its rim sinks into the concavity. (b) Bowl-shaped particle ($u_{mod} = 10.0$) minimally adheres to the vesicle waist. (c) Tetrahedral particle ($u_{mod} = 10.0$) rolls into the vesicle concavity. Images progress from left to right over simulation time.

**Cell-Scale Dynamic Modeling of Membrane Interactions with Arbitrarily Shaped Particles**

Didarul Ahasan Redwan[1], Justin Reicher[2], Xin Yong[1,2]*

[1] Department of Mechanical and Aerospace Engineering, University at Buffalo, Buffalo, NY 14260, USA.

[2] Department of Mechanical Engineering, Binghamton University, Binghamton, NY, 13902

*Email: xinyong@buffalo.edu

## Supplementary Table

Table S1: Computational methods for studying the vesicle–particle interaction system.

| Methods | Capabilities | Limitations |
|---|---|---|
| Triangulated surface (force-based)[1-6] | • Explicit mesh representation<br>• Dynamic evolution of vesicle shape<br>• Easily add more Physics<br>• Large deformations | • High computational cost<br>• Stiff constraints<br>• No topological change<br>• No built-in hydrodynamics |
| Triangulated surface (Monte Carlo)[7-14] | • Equilibrium sampling<br>• Thermal fluctuations<br>• Shape flexibility<br>• Multi-energy terms | • No real-time dynamics<br>• Slow convergence<br>• No hydrodynamics<br>• Limited topology moves |
| Phase-field/level-set (diffuse interface)[15-19] | • Topology changes<br>• Multicomponent fields<br>• Fluid coupling<br>• Large deformations | • High grid cost<br>• Interface thickness<br>• Blurred locality |
| Continuum theory (analytical)[20-23] | • Closed-form criteria<br>• Symmetric solutions<br>• Phase diagrams<br>• Benchmarking | • Fixed topology<br>• Hard to add complex physics<br>• Near-contact issues |

## Supplementary Figures

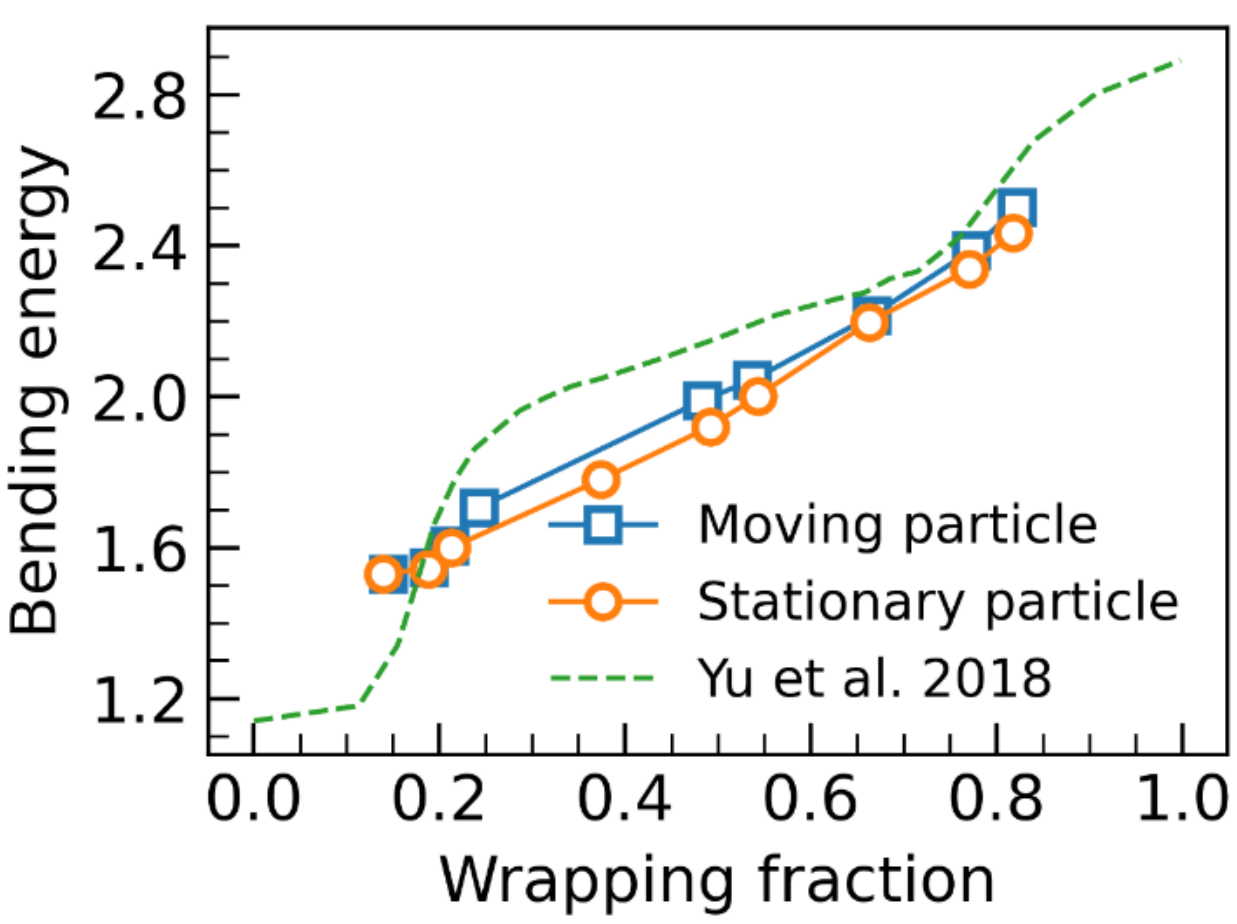

Figure S1: Comparison of normalized bending energy for varying wrapping fraction for a cubical particle interacting with a biconcave vesicle. Simulations were performed with a rescaled adhesion strength of $U_{mod} = 2.0$, a spontaneous curvature of $H_0 = 0.032/R_p$, and a volume constraint of $k_v = 2.0$, targeting a reduced vesicle volume of 0.65 to preserve the biconcave shape. Green dashed lines represent the reference values extracted from Ref. 24. Orange and blue markers correspond to simulations with stationary and mobile particles, respectively.

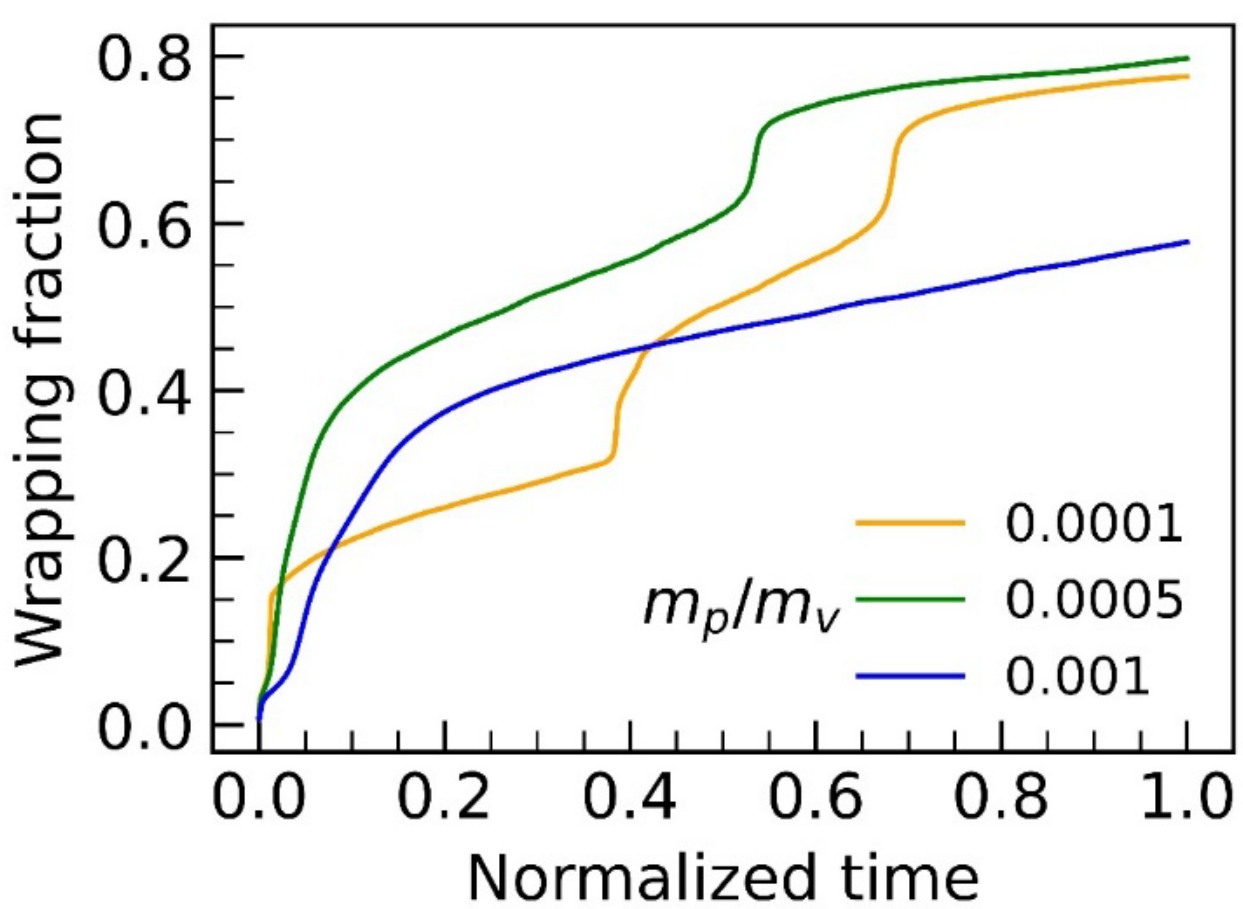

Figure S2: Mass-dependent wrapping progression of cubical particles interacting with a biconcave vesicle. Time evolution of the wrapping fraction for particles with varying particle-to-vesicle mass ratios $m_p/m_v$ = 0.0001 (yellow), 0.0005 (green), and 0.001 (blue).

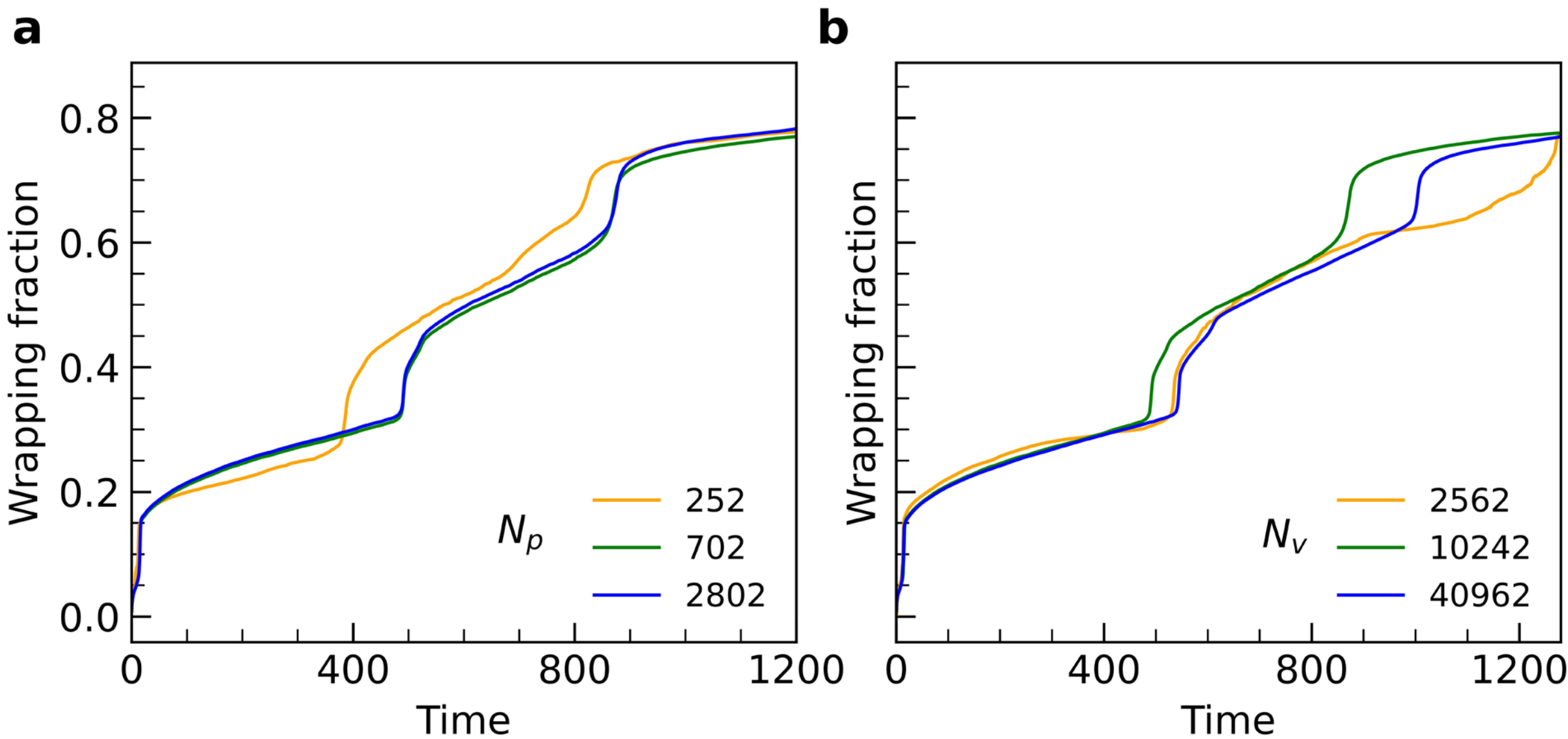

Figure S3: Wrapping fraction as a function of time for cubical particles and biconcave vesicles resolved with different mesh densities. (a) Particles discretized into $N_p = 252$ (orange), 702 (green), and 2802 (blue) vertices interact with a vesicle with $N_v = 10242$ vertices. (b) Vesicles discretized into $N_v = 2556$ (orange), 10242 (green), and 40962 (blue) vertices interact with a particle of $N_p = 702$ vertices.

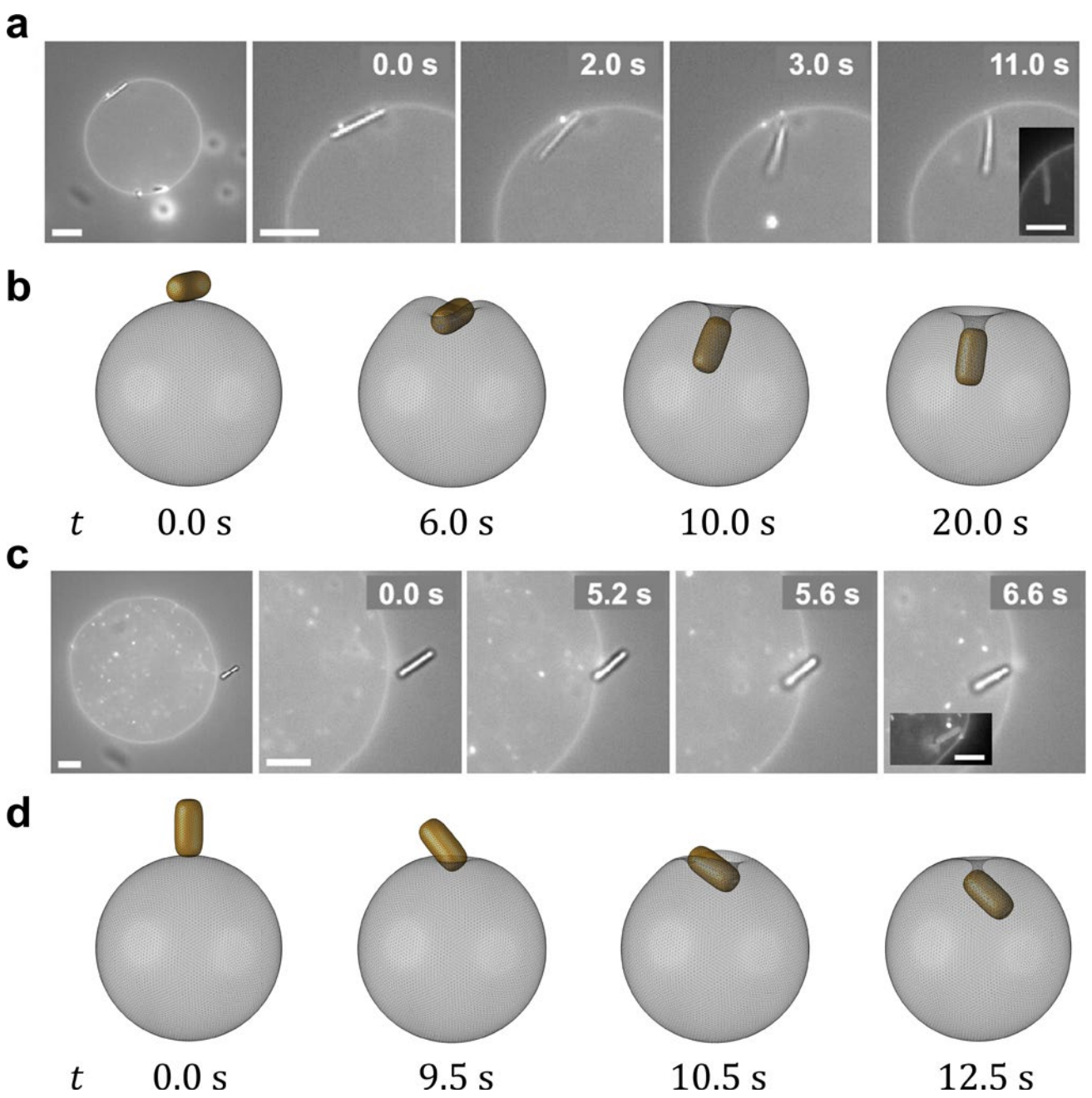

Figure S4: Sequential snapshots of rod-shaped particles engulfed by a spherical vesicle under two initial orientations: (a, b) side-wise and (c, d) tip-wise contact. The experimental snapshots (a, c) are reproduced from Ref. 25.

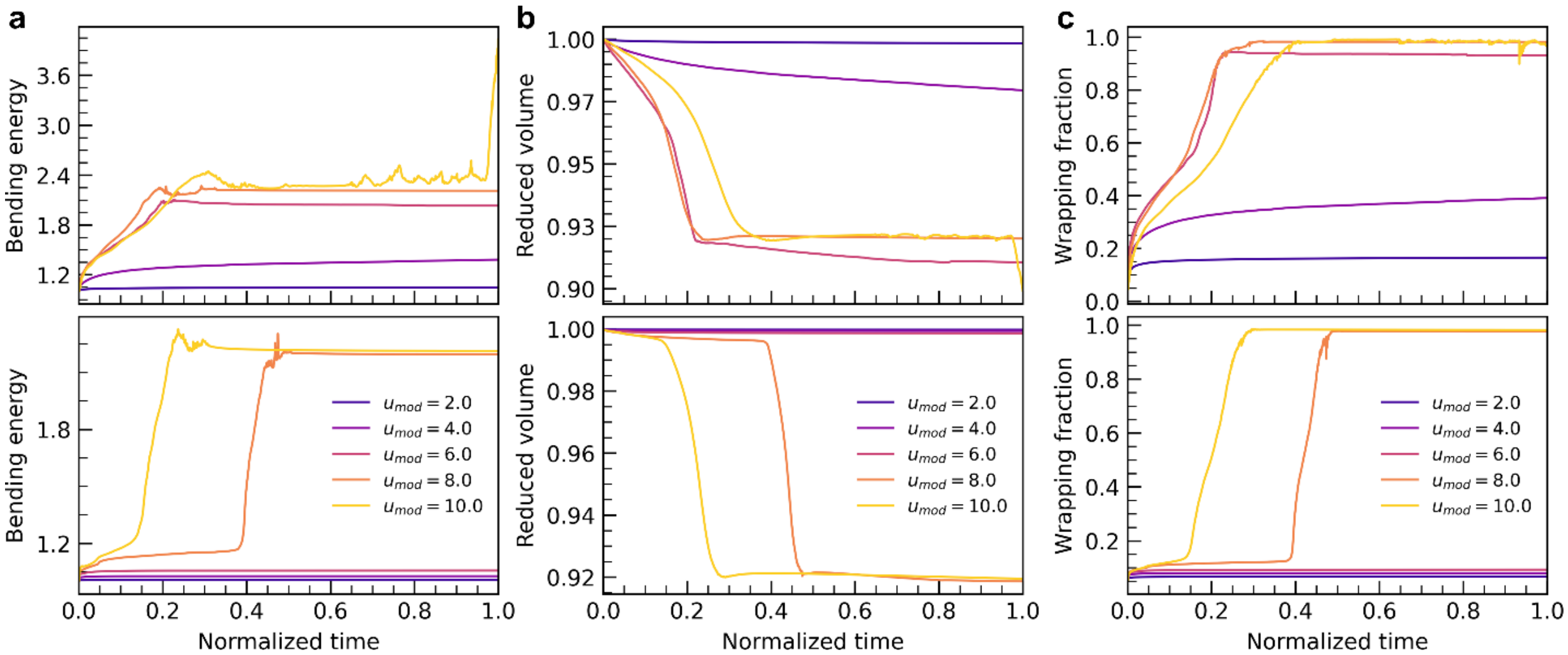

Figure S5: Membrane deformation dynamics for rod wrapping under two initial orientations. Time evolution of (a) membrane bending energy and (b) vesicle reduced volume and (c) wrapping fraction for side-wise contact (top panels) and tip-wise contact (bottom panels). Curves correspond to rescaled adhesion strengths $u_{mod} = 2.0 - 10.0$. Time is normalized by the duration required to reach the steady-state wrapped configuration.

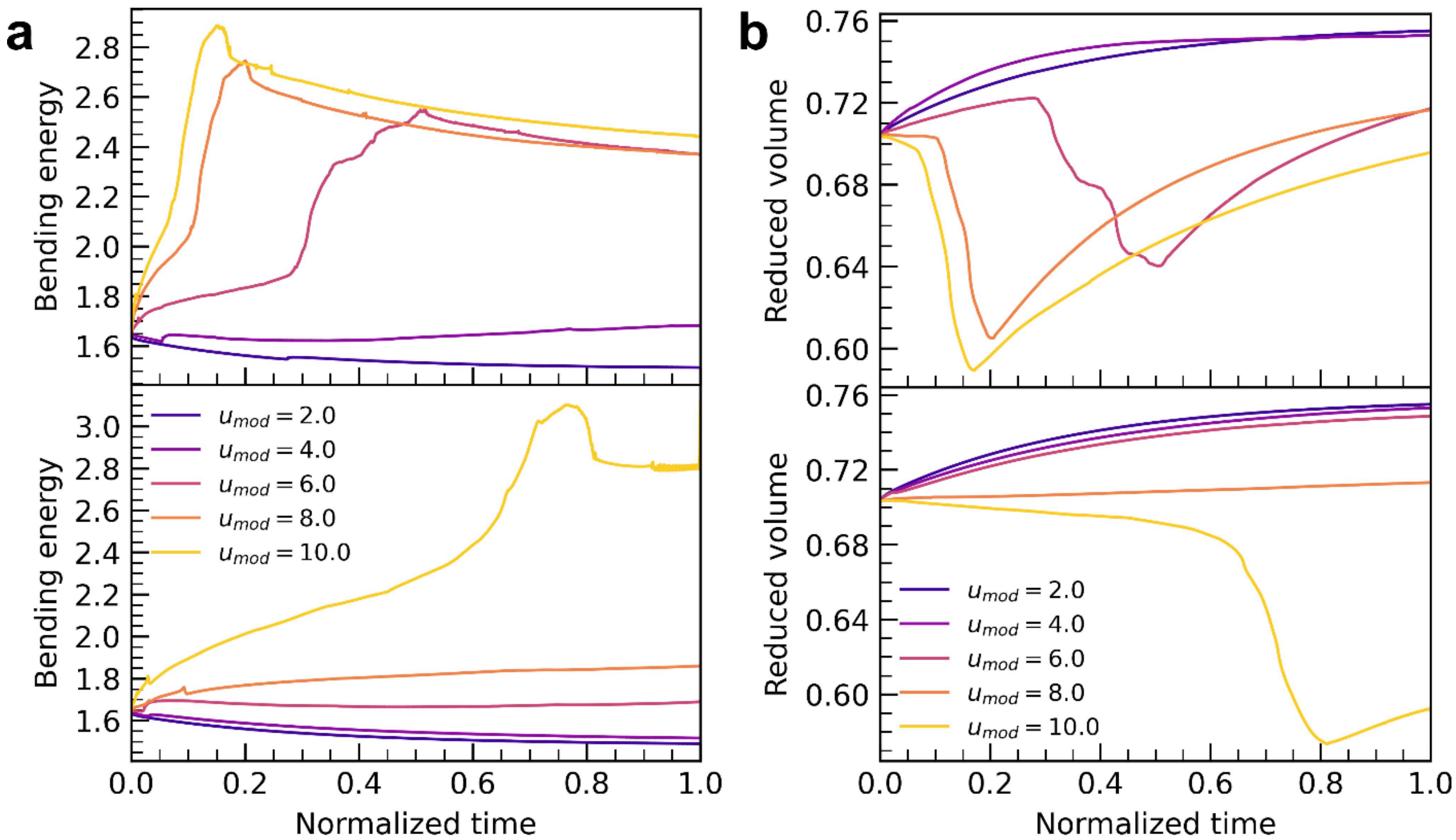

Figure S6: Membrane deformation dynamics during wrapping of a cubical particle from two initial contact configurations: waist-attached (top panels) and pole-attached (bottom panels). The variations in (a) normalized bending energy and (b) reduced vesicle volume as functions of the normalized time are shown. Each curve corresponds to a rescaled adhesion strength $u_{mod}$ ranging from 2.0 to 10.0.

**Supplementary Videos**

Movie S1: Wrapping dynamics of a cubical particle by a biconcave vesicle with a particle-to-vesicle mass ratio of 0.0001.

Movie S2: Wrapping dynamics of a cubical particle by a biconcave vesicle with a particle-to-vesicle mass ratio of 0.0005.

Movie S3: Wrapping dynamics of a cubical particle by a biconcave vesicle with a particle-to-vesicle mass ratio of 0.001.

Movie S4: Wrapping dynamics of a cubical particle initially positioned at the saddle-shaped waist region of a cigar-shaped vesicle.

Movie S5: Wrapping dynamics of a cubical particle initially positioned at the convex north pole of the cigar-shaped vesicle.